\documentclass[aps,prd,twocolumn,groupedaddress,amssymb,eqsecnum,showpacs,
nofootinbib]{revtex4}
\usepackage{graphicx}
\usepackage{bm}

\begin{document}

\preprint{CMU-HEP-04-04}

\title{Fermionic $\alpha$-vacua}

\author{Hael Collins}
\email{hael@physics.umass.edu}
\affiliation{Department of Physics, University of Massachusetts, 
Amherst, MA\ \ 01003}
\affiliation{Department of Physics, Carnegie Mellon University, 
Pittsburgh, PA\ \ 15213}

\date{\today}

\begin{abstract}
A spin ${1\over 2}$ particle propagating in a de Sitter background has a one parameter family of states which transform covariantly under the isometry group of the background.  These states are the fermionic analogues of the $\alpha$-vacua for a scalar field.  We shall show how using a point-source propagator for a fermion in an $\alpha$-state produces divergent perturbative corrections.  These corrections cannot be used to cancel similar divergences arising from scalar fields in bosonic $\alpha$-vacua since they have an incompatible dependence on the external momenta.  The theory can be regularized by modifying the propagator to include an antipodal source.
\end{abstract}


\maketitle

\section{Introduction}
\label{intro}

While de Sitter space shares the same number of isometries as Minkowski space, field theories exhibit some surprising properties in this simplest of curved backgrounds.  An immediate example is the enormous stretching of scales in de Sitter space which naturally connects short distances in the past to large distances today.  This rapid expansion is a familiar and very appealing feature of inflation \cite{textbooks}.  During the slow-roll regime of inflation, for which de Sitter space is an idealization, quantum fluctuations grow exponentially large and eventually seed the large scale structure of the universe.  Depending upon the Hubble scale and the duration of inflation, this large scale structure could have been determined by fluctuations which occurred at or well below the Planck scale.  Most models of inflation produce far more than the necessary sixty $e$-folds necessary to solve the horizon problem and this connection between large scales and potentially Planckian physics has been called the ``transplanckian problem'' of inflation \cite{brandenberger}, although it has recently been viewed as more of an opportunity, since it could allow the observation of physics well beyond experimentally accessible scales, most typically within an order of magnitude or two above the Hubble scale during inflation.  Non-thermal features of the state for the field driving inflation tend to provide a more robust signal of these transplanckian effects \cite{gary,kaloper,transplanck,ulf,cliff,loop,naidu}.

A further difference from Minkowski space is the existence of a much richer family of invariant or covariant states in de Sitter space.  For a free scalar field in a de Sitter background, the states invariant under the $SO(1,4)$ isometry group can be distinguished by a complex parameter $\alpha$, although only real values of $\alpha$ correspond to $CPT$ invariant theories \cite{tagirov,mottola,allen}.  These $\alpha$-vacua are not the lowest energy eigenstates of a globally conserved Hamiltonian, as is the case of the standard Poincar\' e-invariant vacuum of Minkowski space, since de Sitter space does not admit a globally defined time-like Killing vector.  Nevertheless, a unique element in this infinite family, the Bunch-Davies vacuum \cite{bunch}, can be selected by demanding that at short distances or as the curvature of the de Sitter space is taken to zero the state should match with the vacuum of Minkowski space.

Both of these features emphasize the need for understanding quantum field theory---particularly the ideas of decoupling and renormalization---in an expanding background starting from a non-standard state.  For this purpose the $\alpha$-vacua provide an ideal test case to study how these ideas are to be modified in such a setting since the high amount of symmetry of these non-thermal states allows them to be readily analyzed analytically.  It was recently realized that for a scalar field in an $\alpha$-state, a point source propagator does not produce a well behaved perturbation theory \cite{einhorn1,banks,fate}.  One method for expressing this pathological behavior is to impose a cutoff $\Lambda$ on physical three-momentum of the theory.  Loop processes then diverge as $\Lambda\to\infty$ in such a way as cannot be cancelled by simple counterterms.  For example, the one-loop correction to the self-energy in a $\phi^3$ theory diverges linearly with $\Lambda$ and the dependence of this divergent term on the external momentum does not match that of a $\phi^2$ counterterm \cite{fate}.  The resolution of these divergences came with the realization that the propagator should be modified for these states to be the Green's function for {\it two\/} point sources \cite{einhorn2,lowe,taming}.

In this article, we examine the structure and the properties of a spin  ${1\over 2}$ fermion in a de Sitter background, which possesses its own one-parameter set of covariant states \cite{alj,death,deBoer}.  One reason for doing so is to learn whether the double source construction for the scalar field can be circumvented by using fermionic loops to cancel the non-renormalizable divergences from bosonic loops.  While a fermion loop correction to a scalar self-energy also diverges linearly with the cutoff $\Lambda$, here we show that this divergence cannot be cancelled by that of the bosonic loop, even allowing an arbitrary fine-tuning of the relative values of $\alpha$ for the scalar and the fermion fields.

As with a bosonic $\alpha$-vacuum, the peculiar divergences in fermion loops arise from an inconsistency between the single-source propagator and the fermionic $\alpha$-state.  In some sense, the physical setting resembles a field theory where we have imposed boundary conditions along an initial time surface \cite{schalm,initprop}.  There, we must also modify the propagator by adding an image source to obtain a consistent perturbation theory; any new divergences that result from this modification only appear as counterterms restricted to the initial surface.  The bulk theory is unchanged.  In de Sitter space, the inconsistency is also resolved by adding a new source term in the definition of the propagator when in an $\alpha$-state.  The remarkable property of de Sitter space is that there exists a special point, the antipode, at which a source can be placed without breaking the $SO(1,4)$ symmetry properties of the state.  For a fermion, the extra antipodal source entails some additional Dirac structure.

The next section derives the $\alpha$-states for a spin ${1\over 2}$ Dirac field in a de Sitter background.  The $\alpha$-propagator for a point source is developed in Sec.~\ref{pointpropagation}.  We then show in Sec.~\ref{fermiloop} that a theory with a Yukawa coupling produces divergences in the one-loop corrections to the scalar propagator which cannot be cancelled by adding simple counterterms to the Lagrangian nor do they cancel divergences from analogous graphs where a scalar loop replaces the fermion.  Section \ref{antiprop} shows that the these divergences can be avoided by modifying the propagator, adding an additional source at the antipode, resulting in a renormalizable theory.  The final section concludes with comments on the relation between $\alpha$-vacua and the problem of quantizing a theory with a specified initial state.

\section{Fermions in de Sitter space}
\label{fermions}

The existence of the fermionic $\alpha$-vacua was originally established in \cite{alj}; here we present the structure of these states in a conformally flat coordinate system.  The action for a free massive fermion propagating in a general curved space-time is given by 
\begin{equation}
S = \int d^4x\, \sqrt{-g}\, \bar\psi 
\left[ i e^\mu_{\ a} \gamma^a D_\mu - m \right] \psi
\label{action}
\end{equation}
where the $e^\mu_{\ a}$ is the vierbein defining a locally flat frame,
\begin{equation}
g_{\mu\nu}\, e^\mu_{\ a} e^\nu_{\ b} = \eta_{ab} . 
\label{vierbien}
\end{equation}
Varying this action with respect to the fermion field yields the Dirac equation,
\begin{equation}
\left[ i e^\mu_a \gamma^a D_\mu - m \right] \psi = 0 . 
\label{dirac}
\end{equation}
In a curved background, the covariant derivative includes a term for the spin connection, $\omega_{\mu ab}$,
\begin{equation}
D_\mu = \partial_\mu + {\textstyle{1\over 2}} \omega_{\mu ab} \sigma^{ab}
\label{Ddef}
\end{equation}
where 
\begin{equation}
\sigma^{ab} = {\textstyle{1\over 4}} \left[ \gamma^a,\gamma^b \right] . 
\label{sigmadef}
\end{equation}
In terms of the vierbein, the spin connection is given by  
\begin{eqnarray}
\omega_{\mu ab} 
&=& {\textstyle{1\over 2}} e^\nu_{\ a} 
(\partial_\mu e_{b\nu} - \partial_\nu e_{b\mu})
\nonumber \\
&&
- {\textstyle{1\over 2}} e^\nu_{\ b} 
(\partial_\mu e_{a\nu} - \partial_\nu e_{a\mu})
\nonumber \\
&&
- {\textstyle{1\over 2}} e^\nu_{\ a} e^\lambda_{\ b} 
(\partial_\nu e_{c\lambda} - \partial_\lambda e_{c\nu})
e_\mu^{\ c} . 
\label{spinconnection}
\end{eqnarray}

\begin{figure}[!tbp]
\includegraphics{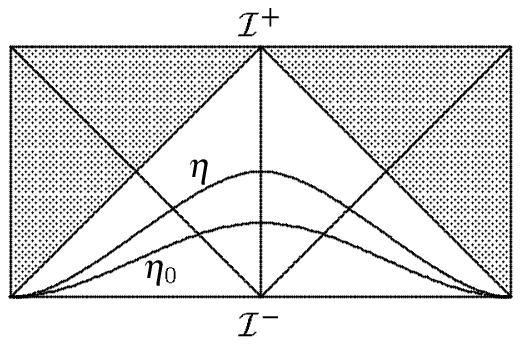}
\caption{The Penrose diagram for de Sitter space.  The conformally flat coordinates of Eq.~(\ref{metric}) cover the unshaded region and surfaces of constant $\eta$ are as shown.\label{penrose}}
\end{figure}

A standard choice for coordinatizing de Sitter space \cite{houches} is provided by writing the metric in a conformally flat form, 
\begin{equation}
ds^2 = g_{\mu\nu}\, dx^\mu dx^\nu 
= {\eta_{\mu\nu}\, dx^\mu dx^\nu\over H^2\eta^2} 
= {d\eta^2 - d\vec x\cdot d\vec x\over H^2\eta^2} , 
\label{metric}
\end{equation}
where $\eta \in [-\infty,0]$ and $H$ is the Hubble constant; we shall most often choose our units so that $H=1$ except later when comparing analogous fermionic and bosonic loops corrections.  The spatial flatness of these coordinates permits a simple expression for the fields and Green's functions for either a spatial or a momentum representation.  These coordinates are also equivalent to the standard set used in inflation by defining $H\eta = -e^{-Ht}$.  Although they cover only half of de Sitter space, indicated by the unshaded region of Fig.~\ref{penrose}, the complementary patch is covered by an analogous set of coordinates given by taking $\eta\to \eta_A = -\eta$ with $\eta_A\in [\infty,0]$.

In these coordinates, the vierbein becomes (with $H=1$) 
\begin{equation}
e^\mu_{\ a} = \eta\, \delta^\mu_{\ a}
\label{CFvierbien}
\end{equation}
and the spin connection and the covariant derivative of a spin ${1\over 2}$ field are respectively 
\begin{equation}
\omega_{\mu ab} = {1\over\eta} 
\left[ \eta_{\mu b} \delta_a^0 - \eta_{\mu a} \delta_b^0 \right]
\label{CFspinconnection}
\end{equation}
and 
\begin{equation}
D_\mu = \partial_\mu + {1\over 4\eta} \eta_{\mu a} 
\left[ \gamma^0, \gamma^a \right] , 
\label{CFD}
\end{equation}
so that the Dirac equation is 
\begin{equation}
i\gamma^0 \bigl[ \eta \partial_\eta - {\textstyle{3\over 2}} \bigr] \psi
+ i \eta \vec\gamma \cdot \vec\nabla \psi - m \psi = 0 . 
\label{confDirac}
\end{equation}
Since the metric is spatially flat, a general fermion $\psi(x)$ is conveniently expanded in terms of positive and negative frequency modes, 
\begin{eqnarray}
&&\!\!\!\!\!\!\!\!\!\!\psi(x) = 
\label{fermion} \\
&&\!\!\!\!\!\sum_{s=\pm 1} \int {d^3\vec k\over (2\pi)^3}\, 
\left[ 
u_{\vec k}^{(s)}(\eta) e^{i\vec k\cdot \vec x}\, b_{\vec k}^{(s)}
+ v_{\vec k}^{(s)}(\eta) e^{-i\vec k\cdot \vec x}\, c_{\vec k}^{(s)\dagger}
\right]
\nonumber
\end{eqnarray}
where the creation and annihilation operators obey the following anticommutation relations
\begin{eqnarray}
\left\{ b_{\vec k}^{(r)}, b_{\vec k'}^{(s)\dagger} \right\}
&=& (2\pi)^3 \delta^{rs} \delta^3(\vec k-\vec k')
\nonumber \\
\left\{ c_{\vec k}^{(r)}, c_{\vec k'}^{(s)\dagger} \right\}
&=& (2\pi)^3 \delta^{rs} \delta^3(\vec k-\vec k')
\label{commsforcbs}
\end{eqnarray}
All other anticommutators vanish.

The Dirac equation can be expressed either as two coupled first order differential equations in terms of a pair of two-component spinors or as two uncoupled second order equations.  Of the two constants of integration, one is fixed the canonical equal-time anticommutation relation.  The spinor field $\psi$ and its conjugate momentum, 
\begin{equation}
\pi = {i\over\eta^3} \psi^\dagger , 
\label{conjmom}
\end{equation}
satisfy the relation,
\begin{equation}
\left\{ \psi^A(\eta,\vec x), \pi_B(\eta,\vec y) \right\}
= i \delta^A_B \delta^3(\vec x-\vec y) , 
\label{equalcomm}
\end{equation}
where the indices in this equation refer to the components of the Dirac spinors.  This condition fixes the normalization of the modes.  Inserting the mode expansion of Eq.~(\ref{fermion}) into this relation and applying Eq.~(\ref{commsforcbs}), we find that the components of the modes should satisfy 
\begin{equation}
\sum_s \left[ [u_{\vec k}^{(s)}]^A [u_{\vec k}^{(s)\dagger}]_B 
+ [v_{\vec k}^{(s)}]^A [v_{\vec k}^{(s)\dagger}]_B \right] 
= \eta^3 \delta^A_B . 
\label{Sspin}
\end{equation}
The second constant of integration corresponds to specifying the vacuum state of the fermion.

\subsection{The Bunch-Davies vacuum}

While it is not possible to define a globally conserved energy in de Sitter space, with respect to which the vacuum is the lowest energy state, a standard vacuum can be selected by prescribing that at distances much shorter than the natural curvature length associated with de Sitter space, the mode functions should match with the positive and negative frequency solutions defined for flat space.  This prescription allows a theory in de Sitter space effectively to inherit the renormalizability of the analogous theory in flat space.  This prescription generalizes that used by Bunch and Davies \cite{bunch} to define the vacuum for a scalar field, so we shall also refer to this state as the ``Bunch-Davies vacuum.''

Demanding that the leading time dependence of the modes should satisfy $u_k^{(s)}(\eta) \propto e^{-ikt}$ and $v_k^{(s)}(\eta) \propto e^{ikt}$ at short distances completely fixes the mode functions,
\begin{equation}
u_{\vec k}^{(s)}(\eta) 
= {\sqrt{\pi k}\over 2} e^{m\pi/2} \eta^2 e^{-i\theta/2} 
\pmatrix{ H_\nu^{(2)}(k\eta)\, \varphi^{(s)}_{\hat k} \cr
is H_{\nu-1}^{(2)}(k\eta)\, \varphi^{(s)}_{\hat k} \cr}
\label{umodes}
\end{equation}
and
\begin{equation}
v_{\vec k}^{(s)}(\eta) 
= -is {\sqrt{\pi k}\over 2} e^{m\pi/2} \eta^2 e^{i\theta/2} 
\pmatrix{ H_{-\nu}^{(1)}(k\eta)\, \chi^{(s)}_{\hat k} \cr
-is H_{1-\nu}^{(1)}(k\eta)\, \chi^{(s)}_{\hat k} \cr} .
\label{vmodes}
\end{equation}
In these expressions $e^{-i\theta/2}$ represents an arbitrary phase.  The normalization of these mode spinors is consistent with Eq.~(\ref{Sspin}) since the Hankel functions satisfy the following identity, 
\begin{equation}
H_\nu^{(2)}(z) H_{1-\nu}^{(1)}(z) + H_{-\nu}^{(1)}(z) H_{\nu-1}^{(2)}(z) 
= - {4i\over\pi z} e^{i\pi\nu} . 
\label{hankels}
\end{equation}
The indices of the Hankel functions are related to the mass of the fermion, 
\begin{equation}
\nu = {\textstyle{1\over 2}} + im . 
\label{nudef}
\end{equation}
The label $s$ refers to the helicity of the mode.  The two-component spinors $\varphi^{(s)}_{\hat k}$ and $\chi^{(s)}_{\hat k}$ are eigenvectors of the helicity operator, 
\begin{eqnarray}
\hat k\cdot\vec\sigma\, \varphi^{(s)}_{\hat k} &=& s\, \varphi^{(s)}_{\hat k} 
\qquad
s = \pm 1
\nonumber\\
\hat k\cdot\vec\sigma\, \chi^{(s)}_{\hat k} &=& -s\, \chi^{(s)}_{\hat k} , 
\label{omegadef}
\end{eqnarray}
and they are related by 
\begin{equation}
\chi^{(s)}_{\hat k} = -i\sigma^2 (\varphi^{(s)}_{\hat k} )^* . 
\label{Pflip}
\end{equation}
Further properties of these two-component spinors are included in the Appendix.

Note that the negative frequency modes are the charge conjugates of the positive frequency modes, 
\begin{equation}
v_{\vec k}^{(s)}(\eta) = C(\bar u_{\vec k}^{(s)}(\eta))^T;
\label{Cconj}
\end{equation}
in the Dirac representation, the charge conjugation operator is given by $C=i\gamma^0\gamma^2$.

\subsection{The MA transform}

We now define a new vacuum $|\alpha\rangle$ which is annihilated by the operators $b_{\vec k}^{\alpha(s)}$ and $c_{\vec k}^{\alpha(s)}$, given by a Bogolubov transformation of the Bunch-Davies operators, 
\begin{eqnarray}
b_{\vec k}^{\alpha(s)} &=& 
N_\alpha \left[ b_{\vec k}^{(s)} - e^{\alpha^*} c_{-\vec k}^{(s)\dagger} \right]
\nonumber \\
c_{\vec k}^{\alpha(s)} &=& 
N_\alpha \left[ c_{\vec k}^{(s)} + e^{\alpha^*} b_{-\vec k}^{(s)\dagger} \right]
\label{MAtransform}
\end{eqnarray}
with  
\begin{equation}
N_\alpha \equiv {1\over\sqrt{1+e^{\alpha+\alpha^*}}} . 
\label{Nanorm}
\end{equation}
The fermionic $\alpha$-vacuum is then defined to be the state such that 
\begin{equation}
b_{\vec k}^{\alpha(s)} |\alpha\rangle = 
c_{\vec k}^{\alpha(s)} |\alpha\rangle = 0 .
\label{alphavac}
\end{equation}
This Bogolubov transformation is the fermionic analogue \cite{alj,death} of the transformation introduced by Mottola \cite{mottola} and Allen \cite{allen} to define the $\alpha$-vacuum for a scalar field.  

The MA transform also induces a transformation of the mode functions, 
\begin{eqnarray}
u_{\vec k}^{\alpha(s)}(\eta) &=& N_\alpha 
\left( u_{\vec k}^{(s)}(\eta) - e^\alpha v_{-\vec k}^{(s)}(\eta) \right)
\nonumber \\
v_{\vec k}^{\alpha(s)}(\eta) &=& N_\alpha 
\left( v_{\vec k}^{(s)}(\eta) + e^{\alpha^*} u_{-\vec k}^{(s)}(\eta) \right) .
\label{MAonmodes}
\end{eqnarray}
In terms of the conformally flat patch, the $\alpha$ modes assume the form
\begin{eqnarray}
u_{\vec k}^{\alpha(s)}(\eta) 
&=& N_\alpha {\sqrt{\pi k}\over 2} e^{m\pi/2} \eta^2 e^{-i\theta/2} 
\label{uAmodes} \\
&&\times
\pmatrix{ 
[H_\nu^{(2)}(k\eta) + is e^\alpha e^{i\theta} H_{-\nu}^{(1)}(k\eta)]\, \varphi_{\hat k}^{(s)} \cr
is [H_{\nu-1}^{(2)}(k\eta) - is e^\alpha e^{i\theta} H_{1-\nu}^{(1)}(k\eta)]\, \varphi_{\hat k}^{(s)} \cr}
\nonumber
\end{eqnarray}
and
\begin{eqnarray}
&&\!\!\!\!\!\!\!\!\!\!
v_{\vec k}^{\alpha(s)}(\eta) 
\label{vAmodes} \\
&&= - is N_\alpha {\sqrt{\pi k}\over 2} e^{m\pi/2} \eta^2 e^{i\theta/2} 
\nonumber \\
&&\quad\times
\pmatrix{ 
[H_{-\nu}^{(1)}(k\eta) + is e^{\alpha^*} e^{-i\theta} H_\nu^{(2)}(k\eta)]\, \chi_{\hat k}^{(s)} \cr
-is [H_{1-\nu}^{(1)}(k\eta) - is e^{\alpha^*} e^{-i\theta} H_{\nu-1}^{(2)}(k\eta)]\, \chi_{\hat k}^{(s)} \cr} , 
\nonumber
\end{eqnarray}
respectively.  Note that the $\alpha$ modes are also charge conjugates of each other, 
\begin{equation}
v_{\vec k}^{\alpha(s)} = C (\bar u_{\vec k}^{\alpha(s)}(\eta) )^T . 
\label{ACconj}
\end{equation}
For simplicity, we include an $\alpha$ index for the mode functions associated with the fermionic $\alpha$ state so that modes functions written without an index always refer to the Bunch-Davies ($\alpha\to -\infty$) state.

\section{Propagation--point sources}
\label{pointpropagation}

The correct prescription for defining the propagator depends on the state being considered.  This dependence is very familiar in systems with boundary conditions; for example, in flat space, if we were to choose Neumann boundary conditions at some initial time (so that time derivatives vanish there), the ordinary free-field propagator is not consistent with these conditions---since the $\Theta$-functions that enforce the time-ordering are not consistent with Neumann boundary conditions.  By adding a fictitious image source at the same position but with the opposite displacement in time as the physical source, we obtain a consistent structure for the time-ordering in the propagator \cite{schalm}.  The extra source encodes the propagation of the initial state information.

In de Sitter space, the enhanced family of $SO(1,4)$ covariant states, which is linked with the existence of antipodal pairs of points, makes the construction of the propagator more subtle than in flat space where only the unique vacuum state transforms consistently with the Poincar\' e invariance of the background.  In this section we show how the most na\"\i ve generalization of the flat space propagator,\footnote{To distinguish the $\alpha$-propagators used in this section from those used in the next, we write the former using script characters, e.g.~${\cal S}_\alpha^F$, while the latter will be written as $S_\alpha^F$.  These latter are the Green's functions associated with two sources.  Green's functions without an explicit $\alpha$ index are those for the Bunch-Davies limit, which is the same function in either case.} 
\begin{equation}
[ie^\mu_a \gamma^a D_\mu - m] {\cal S}^F_\alpha(x,y) = 
{\delta^4(x-y)\over\sqrt{-g(x)}}\, {\bf 1} , 
\label{badFeynman}
\end{equation}
is only appropriate for the Bunch-Davies state.  What emerges when we study the loop corrections in an $\alpha$-state is an inconsistency in this definition.  It is reflected in a dependence on antipodal points on the left side of the equation which is absent from the right side.

We begin by separating the propagator ${\cal S}_\alpha^F(x,y)$ in Eq.~(\ref{badFeynman}) into two-point functions,
\begin{equation}
{\cal S}_\alpha^F(x,y) = \Theta(\eta-\eta')\, {\cal S}_\alpha^>(x,y) 
- \Theta(\eta'-\eta)\, {\cal S}_\alpha^<(x,y) 
\label{feynman}
\end{equation}
where
\begin{eqnarray}
{\cal S}_\alpha^>(x,y) &\equiv& i \langle\alpha | \psi(x)\bar\psi(y) | \alpha\rangle
\label{Greensdef} \\
&=& i \int {d^3\vec k\over (2\pi)^3}\, 
e^{i\vec k\cdot (\vec x-\vec y)}\, 
\sum_s u^{\alpha(s)}_{\vec k}(\eta) \bar u^{\alpha(s)}_{\vec k}(\eta')
\nonumber \\
{\cal S}_\alpha^<(x,y) &\equiv& i \langle\alpha | \bar\psi(y)\psi(x) | \alpha\rangle
\nonumber \\
&=& i \int {d^3\vec k\over (2\pi)^3}\, 
e^{i\vec k\cdot (\vec x-\vec y)}\, 
\sum_s v^{\alpha(s)}_{-\vec k}(\eta) \bar v^{\alpha(s)}_{-\vec k}(\eta') . 
\nonumber 
\end{eqnarray}
with $x=(\eta,\vec x)$ and $y=(\eta',\vec y)$.  The momentum representation of the two-point functions then corresponds to the appropriate sum over products of spinors,
\begin{eqnarray}
{\cal S}_{\alpha,\vec k}^>(\eta,\eta') 
&\!=\!&
i\sum_s u^{\alpha(s)}_{\vec k}(\eta) \bar u^{\alpha(s)}_{\vec k}(\eta')
\nonumber \\
{\cal S}_{\alpha,\vec k}^<(\eta,\eta') 
&\!=\!&
i\sum_s v^{\alpha(s)}_{-\vec k}(\eta) \bar v^{\alpha(s)}_{-\vec k}(\eta') 
\label{FTapropG}
\end{eqnarray}

The sums over the Dirac spinors can be more compactly expressed in terms of the corresponding spin sum for the Bunch-Davies state, with some terms evaluated at antipodal points.  Using the spinor modes in Eqs.~(\ref{umodes}--\ref{vmodes}), we have  
\begin{eqnarray}
&&\!\!\!\!\!\!\!\!\!\!\!\!\!\!\!
S_{\vec k}^>(\eta,\eta') 
\label{bfSdef} \\
&=& {i\pi k\over 4} e^{m\pi} (\eta\eta')^2 
\nonumber \\
&&
\times \left[ \matrix{ 
{\scriptstyle H_\nu^{(2)}(k\eta) H_{1-\nu}^{(1)}(k\eta') \cdot {\bf 1} }
&{\scriptstyle H_\nu^{(2)}(k\eta) H_{-\nu}^{(1)}(k\eta') \cdot i\hat k\cdot\vec\sigma }\cr 
{\scriptstyle H_{\nu-1}^{(2)}(k\eta) H_{1-\nu}^{(1)}(k\eta') \cdot i\hat k\cdot\vec\sigma }
&{\scriptstyle -H_{\nu-1}^{(2)}(k\eta) H_{-\nu}^{(1)}(k\eta') \cdot {\bf 1} }\cr}  \right]
\nonumber \\
&&\!\!\!\!\!\!\!\!\!\!\!\!\!\!\!
S_{\vec k}^<(\eta,\eta')  
\nonumber \\
&=& {i\pi k\over 4} e^{m\pi} (\eta\eta')^2 
\nonumber \\
&&
\times \left[ \matrix{ 
{\scriptstyle H_{-\nu}^{(1)}(k\eta) H_{\nu-1}^{(2)}(k\eta') \cdot {\bf 1} }
&{\scriptstyle - H_{-\nu}^{(1)}(k\eta) H_\nu^{(2)}(k\eta') \cdot i\hat k\cdot\vec\sigma }\cr 
{\scriptstyle - H_{1-\nu}^{(1)}(k\eta) H_{\nu-1}^{(2)}(k\eta') \cdot i\hat k\cdot\vec\sigma }
&{\scriptstyle - H_{1-\nu}^{(1)}(k\eta) H_\nu^{(2)}(k\eta') \cdot {\bf 1} } 
\cr} \right] , 
\nonumber 
\end{eqnarray}
where the lack of an $\alpha$ label indicates the values for the Bunch-Davies limit.  If we note that 
\begin{eqnarray}
H_\nu^{(2)}(z) &\!=\!& - H_{-\nu}^{(1)}(-z) 
\nonumber \\
H_\nu^{(1)}(z) &\!=\!& - H_{-\nu}^{(2)}(-z)
\label{Hankeltrans}
\end{eqnarray}
and define the following Dirac operator 
\begin{equation}
{\cal M} = i \hat k\cdot \vec\gamma \gamma_5 
= \left[ 
\matrix{ i\hat k\cdot\vec\sigma &0 \cr 0 &-i \hat k\cdot\vec\sigma\cr}
\right] , 
\label{calMdef}
\end{equation}
then we can formally write the Dirac structure of the ${\cal S}^>_\alpha(x,y)$ part of the propagator as  
\begin{eqnarray}
{\cal S}_{\alpha,\vec k}^>(\eta,\eta')
&\!=\!& N_\alpha^2 \Bigl\{ 
S_{\vec k}^>(\eta,\eta')
+ e^{\alpha+\alpha^*} {\cal M} S_{\vec k}^>(-\eta,-\eta') {\cal M}^\dagger 
\nonumber \\
&&\qquad
- e^\alpha e^{i\theta} {\cal M} S_{\vec k}^>(-\eta,\eta') 
\nonumber \\
&&\qquad
- e^{\alpha^*} e^{-i\theta} S_{\vec k}^>(\eta,-\eta') {\cal M}^\dagger
\Bigr\}
\label{ApropDiracG} 
\end{eqnarray}
while that of the $S^<_\alpha(x,y)$ part is 
\begin{eqnarray}
{\cal S}_{\alpha,\vec k}^<(\eta,\eta') 
&\!=\!& N_\alpha^2 \Bigl\{ 
S_{\vec k}^<(\eta,\eta') 
+ e^{\alpha+\alpha^*} {\cal M} 
S_{\vec k}^<(-\eta,-\eta') {\cal M}^\dagger
\nonumber \\
&&\qquad
- e^{\alpha^*} e^{-i\theta} {\cal M} S_{\vec k}^<(-\eta,\eta') 
\nonumber \\
&&\qquad
- e^\alpha e^{i\theta} S_{\vec k}^<(\eta,-\eta') {\cal M}^\dagger 
\Bigr\} . 
\label{ApropDiracL} 
\end{eqnarray}

Recall that in conformally flat coordinates the antipodes associated with points $x$ and $y$ can be formally written as $x_A=(-\eta,\vec x)$ and $y_A=(-\eta',\vec y)$.  By defining the Fourier transform of the operator ${\cal M}$ to be $\tilde{\cal M}$, we can write the position-space representation of the two-point functions as 
\begin{eqnarray}
&&\!\!\!\!\!\!\!\!\!\!\!\!\!\!\!\!
{\cal S}_\alpha^>(x,y) =
\label{Aprop} \\
&&
N_\alpha^2 \Bigl\{
S_E^>(x,y) 
+ e^{\alpha+\alpha^*} \tilde{\cal M} S_E^>(x_A,y_A) \tilde{\cal M}^\dagger
\nonumber \\
&&
- e^\alpha e^{i\theta} \tilde{\cal M} S_E^>(x_A,y) 
%\nonumber \\
%&&
- e^{\alpha^*} e^{-i\theta} S_E^>(x,y_A) \tilde{\cal M}^\dagger \Bigr\}
\nonumber \\
&&\!\!\!\!\!\!\!\!\!\!\!\!\!\!\!\!
{\cal S}_\alpha^<(x,y) = 
\nonumber \\
&&
N_\alpha^2 \Bigl\{
S_E^<(x,y) 
+ e^{\alpha+\alpha^*} \tilde{\cal M} S_E^<(x_A,y_A) \tilde{\cal M}^\dagger
\nonumber \\
&&
- e^{\alpha^*} e^{-i\theta} \tilde{\cal M} S_E^<(x_A,y) 
%\nonumber \\
%&&
- e^\alpha e^{i\theta} S_E^<(x,y_A) \tilde{\cal M}^\dagger \Bigr\} . 
\nonumber 
\end{eqnarray}
From the expression for ${\cal M}$ in Eq.~(\ref{calMdef}), note that $\tilde{\cal M} = - \tilde{\cal M}^\dagger$ is an anti-hermitian operator.  In these expressions, it is understood that the operator ${\cal M}$ (${\cal M}^\dagger$) acts on the nearest argument of the two-point function to the right (left).

When written in terms of antipodal coordinates, we can begin to see how loop corrections from fermions, based upon the $\alpha$ propagator defined in Eq.~(\ref{feynman}), lead to exactly the same new divergences as arose in the $\alpha$-vacua of a scalar field in de Sitter space.  In the large momentum limit, $k=|\vec k|\to\infty$, the two-point functions simplify considerably since the Hankel functions become proportional to exponentials; for example, the leading behavior of the Bunch-Davies spin sums are given by 
\begin{equation}
\sum_s u^{(s)}_{\vec k}(\eta) \bar u^{(s)}_{\vec k}(\eta') 
\to {(\eta\eta')^{3/2}\over 2k} e^{-ik(\eta-\eta')} 
[k\gamma^0 - \vec k\cdot\vec\gamma] 
\label{UVusum}
\end{equation}
and 
\begin{equation}
\sum_s v^{(s)}_{-\vec k}(\eta) \bar v^{(s)}_{-\vec k}(\eta') 
\to {(\eta\eta')^{3/2}\over 2k} e^{ik(\eta-\eta')} 
[k\gamma^0 + \vec k\cdot\vec\gamma] . 
\label{UVvsum}
\end{equation}
Therefore, the Fourier components of the ${\cal S}_\alpha^>(x,y)$ two-point function in the high momentum limit reduce to  
\begin{eqnarray}
&&\!\!\!\!\!\!\!\!\!\!\!\!\!\!
{\cal S}_{\alpha,\vec k}^>(\eta,\eta') \to 
\label{ApropFUVG} \\
&&
i N_\alpha^2 {(\eta\eta')^{3/2}\over 2k} \Bigl\{ 
e^{-ik(\eta-\eta')} [k\gamma^0 - \vec k\cdot\vec\gamma]
\nonumber \\
&&\qquad
+ e^{\alpha+\alpha^*} e^{ik(\eta-\eta')} 
\gamma_5\gamma^0 [k\gamma^0 - \vec k\cdot\vec\gamma] \gamma_5\gamma^0 
\nonumber \\
&&\qquad
- e^\alpha e^{i\theta} e^{ik(\eta+\eta')} 
\gamma_5\gamma^0 [k\gamma^0 - \vec k\cdot\vec\gamma] 
\nonumber \\
&&\qquad
- e^{\alpha^*} e^{-i\theta} e^{-ik(\eta+\eta')} 
[k\gamma^0 - \vec k\cdot\vec\gamma] \gamma_5\gamma^0 
\Bigr\} 
\nonumber
\end{eqnarray}
while for the ${\cal S}^<_\alpha(x,y)$ two-point function, 
\begin{eqnarray}
&&\!\!\!\!\!\!\!\!\!\!\!\!\!\!
{\cal S}_{\alpha,\vec k}^<(\eta,\eta') \to 
\label{ApropFUVL} \\
&&
i N_\alpha^2 {(\eta\eta')^{3/2}\over 2k} \Bigl\{ 
e^{ik(\eta-\eta')} [k\gamma^0 + \vec k\cdot\vec\gamma]
\nonumber \\
&&\qquad
+ e^{\alpha+\alpha^*} e^{-ik(\eta-\eta')} 
\gamma_5\gamma^0 [k\gamma^0 + \vec k\cdot\vec\gamma] \gamma_5\gamma^0 
\nonumber \\
&&\qquad
- e^{\alpha^*} e^{-i\theta} e^{-ik(\eta+\eta')} 
\gamma_5\gamma^0 [k\gamma^0 + \vec k\cdot\vec\gamma] 
\nonumber \\
&&\qquad
- e^\alpha e^{i\theta} e^{ik(\eta+\eta')} 
[k\gamma^0 + \vec k\cdot\vec\gamma] \gamma_5\gamma^0 
\Bigr\} . 
\nonumber 
\end{eqnarray}
Different signs appear in the $[k\gamma^0 \pm \vec k\cdot\vec\gamma]$ factors since we have chosen the same sign for the three momentum in Eq.~(\ref{Greensdef}) in both cases when defining the Fourier modes.  

In the Bunch-Davies limit, the $\Theta$-functions in the propagator keep the momentum dependence in the exponentials from cancelling between different lines in a loop.  For example, in a loop consisting of two propagators through which momenta $\vec p$ and $\vec p-\vec k$ flow respectively, the $\Theta$-functions only allow the products 
$S_{\vec p}^>(\eta,\eta') S_{\vec p-\vec k}^<(\eta',\eta)$ and $S_{\vec p}^<(\eta,\eta') S_{\vec p-\vec k}^>(\eta',\eta)$ to appear.  From Eqs.~(\ref{ApropFUVG}--\ref{ApropFUVL}), these products are proportional respectively to $e^{-ip(\eta-\eta')} e^{-i|\vec p-\vec k|(\eta-\eta')}$ and $e^{ip(\eta-\eta')} e^{i|\vec p-\vec k|(\eta-\eta')}$ in the $p=|\vec p|\to\infty$ limit.  Since the $p$-dependent parts of the phases do not cancel, it is possible to define a consistent $i\epsilon$ prescription that renders the integral over $p$ finite.  

The appearance of all possible phases in the $\alpha$ case immediately indicates that the time ordering inherited from the Bunch-Davies limit will always produce some phase cancellation.  As an example, the product ${\cal S}_{\alpha,\vec p}^>(\eta,\eta') {\cal S}_{\alpha,\vec p-\vec k}^<(\eta',\eta)$ also occurs in the $\alpha$ case, since the $\Theta$-function structure is identical; but in this product now occur terms of the proportional to $e^{\alpha+\alpha^*} e^{-ip(\eta-\eta')} e^{i|\vec p-\vec k|(\eta-\eta')}$, $e^{\alpha+\alpha^*} e^{ip(\eta+\eta')} e^{-i|\vec p-\vec k|(\eta+\eta')}$ and their complex conjugates.  In the $p\to\infty$ limit, the $p$-dependent part of these phases cancels so that the loop integral will divergence if sufficiently many powers of $p$ appear in the loop integrand.  Stated differently, these loops contain pinched singularities \cite{einhorn1}.

This phase cancellation is a first indication that in using a single $\delta$-function source we have not constructed a Green's function which is compatible with the underlying structure of the state.  We next establish that the pathologies which arise when we use this propagator do not fortuitously cancel among various divergent terms in an actual perturbative correction.

\subsection{Loop corrections from fermions}
\label{fermiloop}

\begin{figure}[!tbp]
\includegraphics{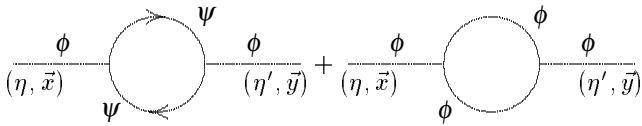}
\caption{The leading loop corrections to a scalar propagator from a theory with a Yukawa coupling to a fermion, $-g\phi\bar\psi\psi$, and a cubic self-coupling, $-{1\over 6}\lambda \phi^3$. 
\label{fbloop}}
\end{figure}

To understand these pathologies better, we examine a simple one-loop graph in a theory with a scalar and a spinor interacting through a Yukawa coupling,
\begin{equation}
{\cal L} = \bar\psi [ie^\mu_{\ a} \gamma^a D_\mu - m] \psi 
+ {\textstyle{1\over 2}} \partial_\mu\phi\partial^\mu\phi 
- {\textstyle{1\over 2}} \mu^2\phi^2 
- g \phi\bar\psi\psi + \cdots . 
\label{yukcoup}
\end{equation}
Later we include as well a scalar self-interaction term,
\begin{equation}
{\cal L}_{\rm int} = \cdots - {\textstyle{1\over 6}} \lambda\phi^3 , 
\label{cubecoup}
\end{equation}
to be able to compare the corrections from a fermion and a scalar loop to the self-energy of the scalar field, as shown in Fig.~\ref{fbloop}.

Since de Sitter space lacks a well-defined $S$-matrix \cite{smatrix}, we evaluate the self-energy corrections to the scalar using the Schwinger-Keldysh formalism \cite{schwinger,keldysh,kt} which essentially corresponds to time-evolving both the ``in'' and ``out'' states of a matrix element.  This evolution is accomplished by formally doubling the interactions of the theory, writing the interacting part of the Hamiltonian as 
\begin{equation}
H_I = g \int_{\eta_0}^\infty {d^3\vec x\over\eta^4}\, \left[
\phi^+\bar\psi^+\psi^+ - \phi^-\bar\psi^-\psi^- \right] , 
\label{Hyukcoup}
\end{equation}
with an analogous field doubling for the scalar self-interaction.  In essence the first term represents the effect of time-evolving the ``in'' state while the second results from time-evolving the ``out'' state.  We assume the system is initially in an general $\alpha$-state, 
\begin{equation}
|{\rm in}(\eta_0)\rangle =|{\rm out}(\eta_0)\rangle 
= |\tilde\alpha, \alpha\rangle , 
\label{inout}
\end{equation}
where $\tilde\alpha$ is the parameter for the bosonic vacuum and $\alpha$ is the parameter for the fermionic state.  When evaluating the contraction of two fields, there are four possibilities for each field depending upon whether the states are labeled with a $\pm$.  The important feature of this formal field doubling is that the time-ordering is such that the arguments of the $-$ fields always occur after and {\it in the opposite order\/} as those of the $+$ fields.  Thus the four possible contractions of two spinors yield four propagators, 
\begin{eqnarray}
{\cal S}_\alpha^{++}(x,y) &=& \Theta(\eta-\eta')\, {\cal S}_\alpha^>(x,y) 
- \Theta(\eta'-\eta)\, {\cal S}_\alpha^<(x,y) 
\nonumber \\
{\cal S}_\alpha^{--}(x,y) &=& \Theta(\eta'-\eta)\, {\cal S}_\alpha^>(x,y) 
- \Theta(\eta-\eta')\, {\cal S}_\alpha^<(x,y) 
\nonumber \\
{\cal S}_\alpha^{-+}(x,y) &=& {\cal S}_\alpha^>(x,y) 
\nonumber \\
{\cal S}_\alpha^{+-}(x,y) &=& - {\cal S}_\alpha^<(x,y) . 
\label{fermiCTP}
\end{eqnarray}
Notice that the time-ordering of a contraction of two $-$ fields, ${\cal S}^{--}_\alpha$, is the opposite that of a contraction of two $+$ fields, ${\cal S}^{++}_\alpha$.  The structure of the contractions of scalar fields is analogous.  A more extensive discussion of the Schwinger-Keldysh formalism as applied to de Sitter space is contained in \cite{fate,taming}.  There we show that that time-evolution of the matrix element of an operator ${\cal O}$ from initial $|\tilde\alpha,\alpha\rangle$ state at $\eta=\eta_0$ is given by
\begin{equation}
{\langle \tilde\alpha, \alpha | T \bigl( {\cal O}(\eta) 
e^{-i\int_{\eta_0}^\infty d\eta' 
\left[ H_I[\psi^+,\phi^+] - H_I[\psi^-,\phi^-] \right]} \bigr) 
| \tilde\alpha, \alpha \rangle \over 
\langle \tilde\alpha, \alpha | T \bigl(
e^{-i\int_{\eta_0}^\infty d\eta' 
\left[ H_I[\psi^+,\phi^+] - H_I[\psi^-,\phi^-] \right]} \bigr) 
| \tilde\alpha, \alpha \rangle  
} . 
\label{inoutmatrix}
\end{equation}

In a generic fermionic $\alpha$-state, the self-energy correction to the scalar propagator contains some terms which diverge linearly in terms of a cutoff $\Lambda$ imposed on the integral of the loop three-momentum.  If we define the Fourier transform of the fermion loop correction in Fig.~\ref{fbloop} as 
\begin{equation}
-i \int {d^3\vec k\over (2\pi)^3}\, e^{i\vec k\cdot(\vec x-\vec y)} 
\Pi_{\vec k}^f(\eta,\eta') ,
\label{Pidef}
\end{equation}
then the linearly divergent term is 
\begin{widetext}
\begin{eqnarray}
\Pi_{\vec k}^f(\eta,\eta')
&=& {ig^2\over\pi^2} {\Lambda\over kH^2} e^{\alpha+\alpha^*} N_\alpha^4 
\int_{\eta_0}^\eta {d\eta_1\over\eta_1} 
\left[ {\cal G}_{\tilde\alpha,k}^>(\eta,\eta_1) 
- {\cal G}_{\tilde\alpha,k}^<(\eta,\eta_1) \right]
\int_{\eta_0}^{\eta'} {d\eta_2\over\eta_2} 
\left[ {\cal G}_{\tilde\alpha,k}^>(\eta',\eta_2) 
- {\cal G}_{\tilde\alpha,k}^<(\eta',\eta_2) \right]
\label{SEfermion} \\
&&\qquad\quad
\times\biggl[ 
{\sin k(\eta_1-\eta_2) - k(\eta_1-\eta_2)\cos k(\eta_1-\eta_2) 
\over (\eta_1-\eta_2)^3 }
+ {\sin k(\eta_1+\eta_2) - k(\eta_1+\eta_2)\cos k(\eta_1+\eta_2) 
\over (\eta_1+\eta_2)^3 }
\biggr]
+ \cdots . 
\nonumber 
\end{eqnarray}
\end{widetext}
The functions ${\cal G}^{>,<}_{\tilde\alpha,k}$ are the Fourier transforms of the Wightman functions for the scalar field,
\begin{equation}
{\cal G}^>_{\tilde\alpha}(x,y) = {\cal G}^<_{\tilde\alpha}(y,x) 
= \langle\tilde\alpha | \phi(x)\phi(y) | \tilde\alpha\rangle . 
\label{Gwight}
\end{equation}
Here we have only explicitly written the new divergent part for the $\alpha$-vacuum although the full self-energy correction contains other finite terms as well as some divergent terms which can be cancelled by a mass counterterm.  These latter terms survive in the Bunch-Davies limit and correspond to the usual need to renormalize the theory, which occurs even in flat space.

An unrenormalized interacting field theory in flat space typically has divergent corrections.  The theory remains predictive since these divergences can be removed by defining rescaled fields and couplings, in terms of which the perturbative corrections are small and finite for small couplings.  This process is possible since the divergent parts of diagrams come from the region, in position space, where a loop shrinks to a point.  Thus such divergences are cancelled with local counterterms.

The divergences in the $\alpha$-states are quite different.  If we shrink the loop in Fig.~\ref{fbloop} to a point, it might appear that the divergence in Eq.~(\ref{SEfermion}) could be cancelled by the insertion of a mass counterterm, 
\begin{equation}
{\cal L}_{c.t.} = - {\textstyle{1\over 2}} \delta m^2 \phi^2 + \cdots .
\label{massct}
\end{equation}
However, the resulting leading correction to the self-energy from a mass counterterm yields instead 
\begin{widetext}
\begin{eqnarray}
\Pi_{\vec k}^{\rm c.t.}(\eta,\eta')
&=& - \delta m^2 \biggl\{ 
\int_{\eta_0}^{\min(\eta,\eta')} {d\eta_1\over\eta_1^4} 
\left[ {\cal G}_{\tilde\alpha,k}^>(\eta,\eta_1) 
- {\cal G}_{\tilde\alpha,k}^<(\eta,\eta_1) \right]
\left[ {\cal G}_{\tilde\alpha,k}^>(\eta',\eta_1) 
- {\cal G}_{\tilde\alpha,k}^<(\eta',\eta_1) \right]
\label{SEdm} \\
&& 
+ \int_{\eta_0}^\eta {d\eta_1\over\eta_1^4} 
\left[ {\cal G}_{\tilde\alpha,k}^>(\eta,\eta_1) 
- {\cal G}_{\tilde\alpha,k}^<(\eta,\eta_1) \right]
{\cal G}_{\tilde\alpha,k}^<(\eta',\eta_1) 
+
\int_{\eta_0}^{\eta'} {d\eta_1\over\eta_1^4} 
{\cal G}_{\tilde\alpha,k}^<(\eta,\eta_1) 
\left[ {\cal G}_{\tilde\alpha,k}^>(\eta',\eta_1) 
- {\cal G}_{\tilde\alpha,k}^<(\eta',\eta_1) \right] 
\biggr\} . 
\nonumber 
\end{eqnarray}
%\end{widetext}
%
The important feature of this contribution is that its dependence on the external momentum, $k = |\vec k|$, differs from that of the divergence from the fermion loop.  This difference means that there is no constant choice for $\delta m^2$ which removes the divergence.  

For comparison, the loop correction from a cubic scalar vertex ${1\over 6}\lambda \phi^3$ is \cite{fate} 
%
%\begin{widetext}
\begin{eqnarray}
\Pi_{\vec k}^b(\eta,\eta')
&=& - {i\lambda^2\over 8\pi^2} {\Lambda\over kH^4} e^{\tilde\alpha+\tilde\alpha^*} \tilde N_{\tilde\alpha}^4 
\int_{\eta_0}^\eta {d\eta_1\over\eta_1^2} 
\left[ {\cal G}_{\tilde\alpha,k}^>(\eta,\eta_1) 
- {\cal G}_{\tilde\alpha,k}^<(\eta,\eta_1) \right]
\int_{\eta_0}^{\eta_1} {d\eta_2\over\eta_2^2} 
\left[ {\cal G}_{\tilde\alpha,k}^>(\eta,\eta_2) 
- {\cal G}_{\tilde\alpha,k}^<(\eta,\eta_2) \right]
\nonumber \\
&&\qquad\qquad\qquad\qquad\qquad
\times 
\biggl[ 
{\sin k(\eta_1-\eta_2)\over\eta_1-\eta_2}
+ {\sin k(\eta_1+\eta_2)\over\eta_1+\eta_2}
\biggr]
+ \cdots
\label{SEscalar}
\end{eqnarray}
\end{widetext}
where we have let the scalar field be in an $\tilde\alpha$-vacuum.  $\tilde N_{\tilde\alpha}$ is the bosonic normalization, 
\begin{equation}
\tilde N_{\tilde\alpha} 
\equiv {1\over\sqrt{1-e^{\tilde\alpha+\tilde\alpha^*}}} .  
\label{bosonnorm}
\end{equation}
In order to be able to compare the self-energy corrections from bosonic and fermionic loops, we have restored the Hubble scale $H$ in these expressions.  Here also the $k$-dependence prevents the possibility of using the $\phi$-loop in Eq.~(\ref{SEscalar}) to cancel the divergence from the original $\psi$-loop in Eq.~(\ref{SEfermion}).

\section{Propagation---antipodal sources}
\label{antiprop}

The divergences in the loop corrections for a field in an $\alpha$-state are not renormalizable in the sense that they cannot be cancelled by a counterterm which contributes through an analogous graph with the loop shrunk to a point.  In the case of the standard scalar $\alpha$-vacuum, these diverges are removed, not by modifying the interaction part of the Lagrangian, but rather by altering the propagator.  This approach emerges naturally from a generalized time-ordering prescription \cite{lowe,taming} or by regarding the $\alpha$-states as squeezed states \cite{einhorn2}.  In either case the propagator is given by the sum of two Bunch-Davies propagators,
\begin{equation}
G_{\tilde\alpha}^F(x,y) = A_{\tilde\alpha}\, G_E^F(x,y) 
+ B_{\tilde\alpha}\, G_E^F(x_A,y) , 
\label{doublescalar}
\end{equation}
corresponding to placing sources at $x=y$ and $x_A=y$, weighted appropriately.  Here $G_E^F(x,y)$ is the Bunch-Davies propagator for a scalar field,
\begin{eqnarray}
G_E^F(x,y) &=& \Theta(\eta-\eta') 
\langle E|\phi(\eta,\vec x) \phi(\eta',\vec y) | E\rangle
\nonumber \\
&&
+ \Theta(\eta'-\eta) 
\langle E|\phi(\eta',\vec y) \phi(\eta,\vec x) | E\rangle . 
\label{doublescalarprop}
\end{eqnarray}
The values for $A_{\tilde\alpha}$ and $B_{\tilde\alpha}$ differ slightly among the various prescriptions but they agree for $CPT$-invariant (real $\tilde\alpha$) theories \cite{lowe,einhorn2}.  The removal of the divergent terms does not depend on the detailed form for these coefficients.

In this section, we generalize this prescription to the fermionic $\alpha$-states.  The origin of the divergences of the last section lay essentially in an inconsistency in the non-local features of the two-point functions in Eqs.~(\ref{Aprop}) and the time-ordering prescription given in Eq.~(\ref{feynman}).  The time-ordering was chosen so that the propagator was the Green's function associated with a single point source, which was appropriate for the standard vacuum choice.  However, for a non-standard choice for the initial state, the propagator must be appropriately modified to be compatible with that state.  In this section we describe this modification, which leads to an essentially unique prescription for obtaining a renormalizable theory in an $\alpha$-state.

Consider a propagator of the following form, 
\begin{equation}
S_\alpha^F(x,y) = A_\alpha S_E^F(x,y) + B_\alpha \tilde{\cal M} S_E^F(x_A,y) . 
\label{Afeyndef}
\end{equation}
We shall demonstrate that this propagator yields no new non-renormalizable loop divergences such as occurred in the previous section.  The operator $\tilde{\cal M}$ again corresponds to the Fourier transform of the Dirac operator ${\cal M}$ defined in Eq.~(\ref{calMdef}) and is always understood to act on the Dirac index of the spinor that depends on the antipodal coordinate.  The part of the propagator which depends on the antipode $x_A$ is given explicitly by
\begin{eqnarray}
-i\tilde{\cal M} S_E^F(x_A,y) &\!=\!& 
\Theta(\eta_A-\eta')\, \langle E | \tilde{\cal M}\psi(x_A) \bar\psi(y) |E\rangle 
\nonumber \\
&&
- \Theta(\eta'-\eta_A)\, \langle E | \bar\psi(y) \tilde{\cal M}\psi(x_A) |E\rangle 
\nonumber \\
&&
\label{Afeynanti}
\end{eqnarray}
with 
\begin{eqnarray}
\tilde{\cal M}\psi(x_A) &=& \sum_s \int{d^3\vec k\over(2\pi)^3}\, 
\Bigl[ 
e^{-i\theta} v_{-\vec k}^{(s)}(\eta) e^{i\vec k\cdot\vec x} b_{\vec k}^{(s)}
\nonumber \\
&&\qquad\qquad\qquad
- e^{i\theta} u_{-\vec k}^{(s)}(\eta) e^{-i\vec k\cdot\vec x} 
c_{\vec k}^{(s)\dagger}
\Bigr] \qquad
\label{Monpsi}
\end{eqnarray}
in conformally flat coordinates.  

The structure of this $\alpha$-propagator is essentially unique since the terms $S_E^F(x,y_A)\tilde{\cal M}^\dagger$ and $\tilde{\cal M} S_E^F(x_A,y_A)\tilde{\cal M}^\dagger$ are related to those already present,  
\begin{eqnarray}
S_E^F(x,y_A)\tilde{\cal M}^\dagger &\!=\!& \tilde{\cal M} S_E^F(x_A,y) 
\nonumber \\
\tilde{\cal M} S_E^F(x_A,y_A)\tilde{\cal M}^\dagger &\!=\!& - S_E^F(x,y) .
\label{antipodereduce}
\end{eqnarray}
Here we have used that antipodal times have the opposite ordering,  $\Theta(\eta_A-\eta'_A)=\Theta(\eta'-\eta)$.  Thus, once we have chosen the time-ordering of the individual terms within the $\alpha$ propagator to be consistent with that of the associated Euclidean two-point functions, Eq.~(\ref{Afeyndef}) represents the most general structure.  Note that since $\tilde{\cal M}$ is anti-hermitian, the sign which would have appeared on the right side of the first line in Eq.~(\ref{antipodereduce}) is cancelled.

\subsection{The path integral with two sources}

The free field generating functional,
\begin{equation}
W_0[\xi,\bar\xi] = 
{\int {\cal D}\psi {\cal D}\bar\psi\, 
   e^{i \int d^4x\, \sqrt{-g} \left\{ {\cal L}_0(x) 
+ \bar\xi(x) \Psi(x) + \bar\Psi(x) \xi(x) \right\} }\over 
\int {\cal D}\psi {\cal D}\bar\psi\, e^{i \int d^4x\, \sqrt{-g} {\cal L}_0(x) } },
\label{freegenfunc}
\end{equation}
is constructed so that a functional derivative with respect to each of the sources yields the correct propagator,
\begin{eqnarray}
\Bigl[ -i {\textstyle{\delta\over\delta\bar\xi(x)}} \Bigr]
\Bigl[ i {\textstyle{\delta\over\delta\bar\xi(y)}} \Bigr]
W_0[\xi,\bar\xi] 
\Bigr|_{\xi=\bar\xi=0} 
&=& \langle \alpha | T \bigl( \psi(x) \bar\psi(y) \bigr) | \alpha \rangle
\nonumber \\
&=& -i S_\alpha^F(x,y) . 
\label{SaFfromW}
\end{eqnarray}
Since the propagator in this theory contains both the $x$ and its antipode $x_A$, the source should couple the field at both points which is represented in the generating functional by the spinor field $\Psi(x)$,
\begin{equation}
\Psi(x) \equiv a_\alpha\, \psi(x) + b_\alpha\, \tilde{\cal M}\psi(x_A) , 
\label{Psidef}
\end{equation}
which through a change of coordinates can be written as coupling the spinor $\psi(x)$ to sources $\bar\xi(x)$ and $\bar\xi(x_A)$.  We assume that $a_\alpha$ and $b_\alpha$ are real.  Extracting the propagator from $W_0[\xi,\bar\xi]$, we obtain Eq.~(\ref{Afeyndef}) when
\begin{equation}
A_\alpha = a_\alpha^2 - b_\alpha^2 
\quad\hbox{and}\quad
B_\alpha = 2a_\alpha b_\alpha . 
\label{AsandBs}
\end{equation} 

The free generating functional can also be written in a form that is quadratic in the sources.  If we shift the fermion field by replacing $\psi(x)$ with 
\begin{equation}
\psi(x) - 
\int d^4y\, \sqrt{-g} S_\alpha^F(x,y) 
{a_\alpha\psi(y) - b_\alpha \tilde{\cal M} \psi(y_A)
\over a_\alpha^2 + b_\alpha^2} ,
\label{psishift}
\end{equation}
the generating functional becomes 
\begin{equation}
W_0[\xi,\bar\xi] = e^{-i \int d^4x\, \sqrt{-g}\int d^4y\, \sqrt{-g}\,
  \bar\xi(x) S_\alpha^F(x,y) \xi(y) } . 
\label{freegenfuncXX}
\end{equation}

The generating functional for an interacting theory is given by extending the Lagrangian in the path integral to include interactions,
\begin{equation}
W[\xi,\bar\xi] = 
{\int {\cal D}\psi {\cal D}\bar\psi\, 
   e^{i \int d^4x\, \sqrt{-g} \left\{ {\cal L}(x) 
+ \bar\xi(x) \Psi(x) + \bar\Psi(x) \xi(x) \right\} }\over 
\int {\cal D}\psi {\cal D}\bar\psi\, e^{i \int d^4x\, \sqrt{-g} {\cal L}(x) } } . 
\label{genfunc}
\end{equation}
For illustration, we examine a general Yukawa interaction of the form
\begin{equation}
{\cal L}_{\rm int} = \bar\psi V_1(x) \psi 
+ \left[ \bar\Psi V_2(x) \psi + \hbox{h.c.} \right]
+ \bar\Psi V_3(x) \Psi . 
\label{yukact}
\end{equation}
For generality, we have coupled a complete set of fermion bilinears constructed from the local field $\psi(x)$ and the non-local field $\Psi(x)$ given in Eq.~(\ref{Psidef}) to potentials $V_1(x)$, $V_2(x)$ and $V_3(x)$ which could represent other fields present in the theory and can include a non-trivial Dirac structure as well.  We have included a general set of interactions, including terms which contain an antipodal dependence, since the sources $\xi$ and $\bar\xi$ have already introduced some antipodal non-locality into the theory.  

The fields that appear in the interactions, whether a $\psi(x)$ or $\Psi(x)$ field, determine the corresponding functional derivative required to rewrite $W[\xi,\bar\xi]$ in terms of $W_0[\xi,\bar\xi]$.  Since both $\bar\psi(x)$ and $\bar\psi(x_A)$ couple to the source $\xi(x)$, the functional derivative used to extract a factor of $\bar\psi(x)$ is constructed remove the antipodal term,
\begin{equation}
{\delta\over\delta\Xi(x)} = {1\over a_\alpha^2+b_\alpha^2} \left[
a_\alpha {\delta\over\delta\xi(x)} 
- b_\alpha \tilde{\cal M}^\dagger {\delta\over\delta\xi(x_A)} \right] ; 
\label{Xidef}
\end{equation}
thus, for example, 
\begin{eqnarray}
{\delta\over\delta\Xi(x)} \int d^4y\, \sqrt{-g}\, \bar\Psi(y) \xi(y) 
= - \bar\psi(x) . 
\label{dXieg}
\end{eqnarray}
A simple functional derivative with respect to $\xi(x)$, extracts instead the linear combination $\Psi(x)$,
\begin{eqnarray}
{\delta\over\delta\xi(x)} \int d^4y\, \sqrt{-g}\, \bar\Psi(y) \xi(y) 
= - \bar\Psi(x) . 
\label{dXiegs}
\end{eqnarray}
Therefore, for the Yukawa interactions given in Eq.~(\ref{yukact}) the corresponding generating functional is 
\begin{widetext}
\begin{equation}
W[\xi,\bar\xi] = 
{e^{i\int d^4x\, \sqrt{-g}\, \left\{
\left[ -i {\delta\over\delta\Xi(x)} \right] 
V_1(x) \left[ i {\delta\over\delta\overline\Xi(x)} \right] 
+ \left[ \left[ -i {\delta\over\delta\xi(x)} \right] V_2(x) 
\left[ i {\delta\over\delta\overline\Xi(x)} \right]  + {\rm h.c.} \right]
+ \left[ -i {\delta\over\delta\xi(x)} \right] V_3(x) 
\left[ i {\delta\over\delta\overline\xi(x)} \right] 
\right\} }
W_0[\xi,\bar\xi]\over 
e^{i\int d^4x\, \sqrt{-g}\, \left\{
\left[ -i {\delta\over\delta\Xi(x)} \right] 
V_1(x) \left[ i {\delta\over\delta\overline\Xi(x)} \right] 
+ \left[ \left[ -i {\delta\over\delta\xi(x)} \right] V_2(x) 
\left[ i {\delta\over\delta\overline\Xi(x)} \right]  + {\rm h.c.} \right]
+ \left[ -i {\delta\over\delta\xi(x)} \right] V_3(x) 
\left[ i {\delta\over\delta\overline\xi(x)} \right] 
\right\} }
W_0[\xi,\bar\xi] 
\bigr|_{\xi=\bar\xi=0}
} . 
\label{genfuncderv}
\end{equation}
\end{widetext}

The appearance of the different types of interactions leads to a richer propagator structure depending upon whether the contraction is between an internal or an external vertex and the the type of interaction.  An $n,\bar n$-point function with $n$ external $\psi$ lines and $\bar n$ external $\bar\psi$ lines is given by 
\begin{eqnarray}
&&\!\!\!\!\!\!\!\!\!\!\!\!\!\!\!\!\!
\langle \alpha | T \bigl( \psi(x_1) \cdots \psi(x_n) 
\bar\psi(y_1) \cdots \bar\psi(y_{\bar n}) \bigr) | \alpha \rangle
\nonumber \\ 
&=& 
\prod_{i=1}^n\Bigl[ -i {\textstyle{\delta\over\delta\bar\xi(x_i)}} \Bigr]
\prod_{j=1}^{\bar n}\Bigl[ i {\textstyle{\delta\over\delta\bar\xi(y_j)}} \Bigr]
W_0[\xi,\bar\xi] 
\Bigr|_{\xi=\bar\xi=0} 
\qquad
\label{npoint}
\end{eqnarray}
which also implicitly generalizes the time-ordering used in the $\alpha$-vacuum to products of more than two fields.  Thus an external point is always associated with a ${\delta\over\delta\xi}$ or a ${\delta\over\delta\bar\xi}$ derivative.  The functional derivative for an internal vertex, however, can be either of these as well as ${\delta\over\delta\Xi}$ or ${\delta\over\delta\bar\Xi}$ depending upon the form of the interaction, as indicated in Eq.~(\ref{genfuncderv}).  A theory whose interactions only contain the field $\Psi(x)$ will only contain $\alpha$ propagators, defined in Eq.~(\ref{Afeyndef}).  But a theory with only local interactions, depending solely on $\psi(x)$, or mixed interactions, as in the term coupled to $V_2(x)$ in Eq.~(\ref{yukact}), will contain several sets of propagators---either a purely Bunch-Davies propagator, 
\begin{equation}
-i S_E^F(x,y) 
= \Bigl[ -i {\textstyle{\delta\over\delta\overline\Xi(x)}} \Bigr]
\Bigl[ i {\textstyle{\delta\over\delta\Xi(y)}} \Bigr]
W_0[\xi,\bar\xi] 
\Bigr|_{\xi=\bar\xi=0} , 
\label{BDpropW}
\end{equation}
a mixed propagator,
\begin{eqnarray}
-i S_m^F(x,y) 
&\equiv& -i \Bigl[ a_\alpha\, S_E^F(x,y) + b_\alpha\, \tilde{\cal M} S_E^F(x_A,y) \Bigr]
\nonumber \\
&=& \Bigl[ -i {\textstyle{\delta\over\delta\bar\Xi(x)}} \Bigr]
\Bigl[ i {\textstyle{\delta\over\delta\xi(y)}} \Bigr]
W_0[\xi,\bar\xi] 
\Bigr|_{\xi=\bar\xi=0} 
\nonumber \\
&=& \Bigl[ -i {\textstyle{\delta\over\delta\bar\xi(x)}} \Bigr]
\Bigl[ i {\textstyle{\delta\over\delta\Xi(y)}} \Bigr]
W_0[\xi,\bar\xi] 
\Bigr|_{\xi=\bar\xi=0} \qquad
\label{MpropW}
\end{eqnarray}
or a genuine $\alpha$-propagator, 
\begin{equation}
-i S_\alpha^F(x,y) 
= \Bigl[ -i {\textstyle{\delta\over\delta\overline\xi(x)}} \Bigr]
\Bigl[ i {\textstyle{\delta\over\delta\xi(y)}} \Bigr]
W_0[\xi,\bar\xi] 
\Bigr|_{\xi=\bar\xi=0} ; 
\label{ApropW}
\end{equation}
here the $\alpha$-propagator is that given in Eq.~(\ref{Afeyndef}) and not the propagator corresponding to a single point source, as discussed in Sec.~\ref{pointpropagation}.  

One of the consequences of Eq.~(\ref{BDpropW}) is that in a theory with only local interactions, so that $\Psi(x)$ does not appear in ${\cal L}_{\rm int}$, automatically has the same divergence structure in loop corrections as a theory in the Bunch-Davies vacuum.  Consequently, the counterterms and the renormalization is identical to the usual vacuum state.  The only source of $\alpha$-dependence is that which occurs in the external legs which will be mixed propagators, Eq.~(\ref{MpropW}).  

A more interesting scenario, which more closely resembles the real-time ordering prescription applied to a scalar theory in \cite{lowe}, is to consider a theory constructed only out of the linear combination $\Psi(x)$.  In this case, all of the fermion propagators in a diagram are $\alpha$-propagators and the renormalization differs slightly from that of the Bunch-Davies limit \cite{taming}.  In the next section we shall show that even with the additional antipodal terms in the double source $\alpha$, no phase cancellation occurs in the UV region of the loop momentum which produced the non-renormalizable divergences seen in the previous section.

\subsection{Loop corrections in a theory with antipodal sources}
\label{fermiloopA}

As an example we examine a theory with a non-local Yukawa interaction of the special form, 
\begin{equation}
{\cal L}_{\rm int} = - g \Phi\bar\Psi\Psi . 
\label{yukintA}
\end{equation}
For complete generality, we have coupled the linear combination $\Psi(x)$ which yields only $\alpha$-propagators in the diagrams to a similar linear combination for the scalar field \cite{taming},
\begin{eqnarray}
\Phi(x) &=& {1\over\sqrt{2}} \Bigl[ A_{\tilde\alpha} + \sqrt{A_{\tilde\alpha}^2 - B_{\tilde\alpha}^2} \Bigr]^{1/2}\, \phi(x) 
\nonumber \\
&&
+ {1\over\sqrt{2}} {B_{\tilde\alpha}\over \left[ A_{\tilde\alpha} + \sqrt{A_{\tilde\alpha}^2 - B_{\tilde\alpha}^2} \right]^{1/2} }\, \phi(x_A) , 
\label{scalarlin}
\end{eqnarray}
constructed so that the scalar lines in a graph correspond to the bosonic $\alpha$-propagator of Eq.~(\ref{doublescalar}).  As before, each field can be in a different invariant state, as we have denoted by writing different labels for the scalar, $\tilde\alpha$, and spinor, $\alpha$, states.

\begin{figure}[!tbp]
\includegraphics{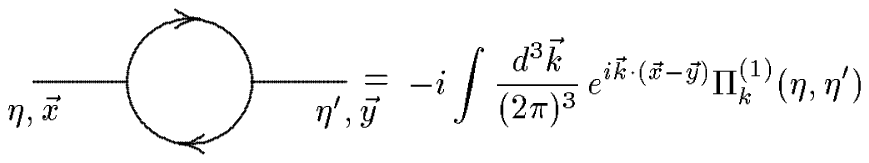}
\caption{The leading loop correction to the scalar propagator from a theory with a special non-local Yukawa coupling to a fermion, $-{1\over 2} g\Phi\bar\Psi\Psi$. 
\label{sig1def}}
\end{figure}

The leading correction to the scalar propagator is given by the one-loop graph shown in Fig.\ref{sig1def}.  
The only new element of for the Schwinger-Keldysh approach is to state how the additional antipodal propagators affect the structure of a contraction between $\psi^\pm$ fermions, 
\begin{eqnarray}
S_{\alpha,\vec k}^{++}(\eta,\eta') 
&=& A_\alpha\, \Theta(\eta-\eta')\, S_{\vec k}^>(\eta,\eta') 
\nonumber \\
&&
- A_\alpha\, \Theta(\eta'-\eta)\, S_{\vec k}^<(\eta,\eta') 
\nonumber \\
&&
+ B_\alpha\, {\cal M} S_{\vec k}^>(-\eta,\eta') 
\nonumber \\
S_{\alpha,\vec k}^{--}(\eta,\eta') 
&=& A_\alpha\, \Theta(\eta'-\eta)\, S_{\vec k}^>(\eta,\eta') 
\nonumber \\
&&
- A_\alpha\, \Theta(\eta-\eta')\, S_{\vec k}^<(\eta,\eta') 
\nonumber \\
&&
- B_\alpha\, S_{\vec k}^<(\eta,-\eta') {\cal M}^\dagger 
\nonumber \\
S_{\alpha,\vec k}^{-+}(\eta,\eta') &=& A_\alpha\, S_{\vec k}^>(\eta,\eta') 
+ B_\alpha\, {\cal M} S_{\vec k}^>(-\eta,\eta') 
\nonumber \\
S_{\alpha,\vec k}^{+-}(\eta,\eta') &=& - A_\alpha\, S_{\vec k}^<(\eta,\eta') 
- B_\alpha\, S_{\vec k}^<(\eta,-\eta') {\cal M}^\dagger  . \qquad 
\label{fermiCTPa}
\end{eqnarray}
Note that the additional terms can be written with either coordinate evaluated at the antipode using the relation
\begin{equation}
{\cal M} S_{\vec k}^>(-\eta,\eta') = 
- S_{\vec k}^<(\eta,-\eta') {\cal M}^\dagger 
\label{switchM}
\end{equation}
which follows directly from Eqs.~(\ref{bfSdef}--\ref{calMdef}).  With these forms for the propagators, we find 
\begin{widetext}
\begin{eqnarray}
\Pi^{(1)}_k(\eta,\eta') 
&=& - ig^2 \int_{\eta_0}^\eta {d\eta_1\over\eta_1^4}
\int_{\eta_0}^{\min(\eta',\eta_1)} {d\eta_2\over\eta_2^4} 
[\tilde G_k^>(\eta,\eta_1) - \tilde G_k^<(\eta,\eta_1)] 
[\tilde G_k^>(\eta',\eta_2) L_k^{{\scriptscriptstyle >,<}}(\eta_1,\eta_2) 
- \tilde G_k^<(\eta',\eta_2) L_k^{{\scriptscriptstyle <,>}}(\eta_1,\eta_2) ]
\nonumber \\
&& 
+ ig^2 
\int_{\eta_0}^{\eta'} {d\eta_2\over\eta_2^4} 
\int_{\eta_0}^{\min(\eta,\eta_2)} {d\eta_1\over\eta_1^4}
[\tilde G_k^<(\eta,\eta_1) L_k^{{\scriptscriptstyle >,<}}(\eta_1,\eta_2) 
- \tilde G_k^>(\eta,\eta_1) L_k^{{\scriptscriptstyle <,>}}(\eta_1,\eta_2) ] 
[\tilde G_k^>(\eta',\eta_2) - \tilde G_k^<(\eta',\eta_2)] 
\nonumber \\
&& 
- ig^2 \int_{\eta'}^\eta {d\eta_1\over\eta_1^4}
\int_{\eta'}^{\eta_1} {d\eta_2\over\eta_2^4} 
[\tilde G_k^>(\eta,\eta_1) - \tilde G_k^<(\eta,\eta_1)] 
\tilde G_k^<(\eta',\eta_2) 
[L_k^{{\scriptscriptstyle >,<}}(\eta_1,\eta_2) 
- L_k^{{\scriptscriptstyle <,>}}(\eta_1,\eta_2) ]
\nonumber \\
&& 
+ ig^2 
\int_\eta^{\eta'} {d\eta_2\over\eta_2^4} 
\int_\eta^{\eta_2} {d\eta_1\over\eta_1^4}
\tilde G_k^<(\eta,\eta_1) 
[\tilde G_k^>(\eta',\eta_2) - \tilde G_k^<(\eta',\eta_2)] 
[L_k^{{\scriptscriptstyle >,<}}(\eta_1,\eta_2)  
- L_k^{{\scriptscriptstyle <,>}}(\eta_1,\eta_2)] . 
\label{loopyukA}
\end{eqnarray}
\end{widetext}
In this expression the fermionic loop integrals have been written as 
\begin{eqnarray}
L_k^{{\scriptscriptstyle >,<}}(\eta_1,\eta_2) 
&=& \int d^3\vec p\, {\rm tr} \bigl[ \tilde S_{\vec p}^>(\eta_1,\eta_2) \tilde S_{\vec p-\vec k}^<(\eta_2,\eta_1) \bigr]
\nonumber \\
L_k^{{\scriptscriptstyle <,>}}(\eta_1,\eta_2) 
&=& \int d^3\vec p\, {\rm tr} \bigl[ \tilde S_{\vec p}^<(\eta_1,\eta_2) \tilde S_{\vec p-\vec k}^>(\eta_2,\eta_1) \bigr] \qquad
\label{loopgtlt}
\end{eqnarray}
and we have grouped together the two-point functions to include both ordinary and antipodal pieces; for the fermions we have  
\begin{eqnarray}
\tilde S_{\vec k}^>(\eta_1,\eta_2) 
&\!\equiv\!& A_\alpha\, S_{\vec k}^>(\eta_1,\eta_2) 
+ B_\alpha\, {\cal M} S_{\vec k}^>(-\eta_1,\eta_2)
\nonumber \\
\tilde S_{\vec k}^<(\eta_1,\eta_2) 
&\!\equiv\!& A_\alpha\, S_{\vec k}^<(\eta_1,\eta_2) 
+ B_\alpha\, S_{\vec k}^<(\eta_1,-\eta_2) {\cal M}^\dagger \qquad
\label{Slincomb}
\end{eqnarray}
while for the scalar field,  
\begin{equation}
\tilde G_k^{>,<}(\eta,\eta') 
\equiv A_{\tilde\alpha}\, G_k^{>,<}(\eta,\eta') 
+ B_{\tilde\alpha}\, G_k^>(-\eta,\eta') . 
\label{Glincomb}
\end{equation}

In the ultraviolet region of the the loop integral, the leading $p$-dependence of either integral in Eq.~(\ref{loopgtlt}) is the same, 
\begin{eqnarray}
&&\!\!\!\!\!\!\!\!\!\!\!\!\!\!\!\!\!\!\!\!\!
{\rm tr} \bigl[ \tilde S_{\vec p}^>(\eta_1,\eta_2) \tilde S_{\vec p-\vec k}^<(\eta_2,\eta_1) \bigr] 
\nonumber \\
&\to& - (\eta_1\eta_2)^3 \bigl[ 1 + \hat p \cdot \widehat{p-k} \bigr] 
\nonumber \\
&&\bigl[ 
A_\alpha^2 e^{-i[p+|\vec p-\vec k|](\eta_1-\eta_2)} 
- B_\alpha^2 e^{i[p+|\vec p-\vec k|](\eta_1+\eta_2)} 
\bigr] 
\nonumber \\
&&+ \cdots
\label{UVStildes}
\end{eqnarray}
Note that there is no longer any phase cancellation in the far UV of the loop integral, such as occurred in the previous section.  In fact, the second term will never present any new divergences since $\eta_1+\eta_2<0$ and it is always possible to define an $i\epsilon$ prescription which renders the $B_\alpha^2$-term finite.

The remaining term, proportional to $A_\alpha^2$, has exactly the same structure as the loop divergence in the Bunch-Davies vacuum.  It can be renormalized therefore in the same way as in the standard vacuum through a wave-function and mass renormalization.

\subsection{Discussion}
\label{discuss}

While the construction of the two-source propagator arose from a need to remove the non-renormalizable divergences that appeared for the single-source propagator, it can be viewed from the more general vantage of the need to write the propagator consistently with the initial state.  This perspective should apply even if the state we are using is well behaved in the ultraviolet---approaching the Bunch-Davies state at short distances---but dramatically different at longer distances, such as the ``truncated $\alpha$-vacua'' \cite{ulf,loop}.  It would be useful to learn whether using the construction here leads to a different estimate for the size of transplanckian effects in the cosmic microwave background.

The existence of fermionic $\alpha$-states can lead to more important role for fermions in inflation than is usually assumed.  For example, when the inflaton is coupled to a fermion in a non-thermal state, the loop corrections can be significantly larger than would be expected from the Bunch-Davies vacuum \cite{loop}.  Moreover, since this effect only depends on the state of the field in the loop, it can occur even when the inflaton is itself in the Bunch-Davies vacuum.  The presence of many potential fermionic degrees of freedom in the standard model allows for further enhancement, depending upon what fraction of the fermions are in a non-thermal state during inflation.

It is important to note that if the $\alpha$-vacua are to be applied to model of inflation, then the model must allow for their decay once the inflationary phase has ended.  Just as with the scalar $\alpha$-vacuum, the value of $\alpha$ today for a fermion is severely constrained by cosmic ray measurements.  The values of $\alpha$ during inflation and during a possible de Sitter phase today, however, do not need to be the same.  The $\alpha$-states are only the vacua of a pure de Sitter background; the de Sitter symmetry is clearly broken in our universe.  Nothing therefore precludes the decay of $\alpha$ to an observationally acceptable value after inflation has ended \cite{ulf,lowecr}.

\section{Conclusions}
\label{conclude}

While interesting in itself, the $\alpha$-vacuum is perhaps even more valuable as an illustration of the new elements required to describe the propagation of a field in a state other than the standard vacuum.  One of these new elements is that the propagator is modified.  In the $\alpha$ case, it becomes the Green's function associated with two sources:  one corresponds to the physical source while the other is located at its antipode.  The antipodal source encodes the effect of the background state on the propagation of a fermion very analogously to how a fictitious image charge encodes the effects of a conducting plate on the propagation of a charge in classical electrodynamics.

Ultimately we would like to extend this formalism to a generic initial state in a general Robertson-Walker background.  The enormous stretching of scales during inflation would suggest that the power spectrum of the cosmic microwave background should be sensitive to the details of a field theory in the deeply ultra-violet region.  However, the observed spectrum is still completely consistent with having the inflaton---and all other fields which could appreciably affect it, such as the fermions discussed here---in the Bunch-Davies vacuum.  What is needed to reconcile these two observations is an extension of the principle of decoupling to rapidly expanding backgrounds.  

In flat space, we have a well defined prescription for understanding the corrections from a more fundamental theory valid at high energies on the effective theories we apply at low energies \cite{eft}.  Similarly, we need to consider the size of the error we make in choosing a Bunch-Davies state when the actual short distance properties of the state are different.  To do so requires being able to renormalize the theory in that state.  By examining the $\alpha$-vacuum, we have learned that the propagator must be modified to obtain the correctly renormalized theory.  More generally, while the $\alpha$-states respect the background space-time symmetries, an arbitrary initial state does not.  This lessened symmetry means that we must also renormalize the initial state \cite{initprop,schalm}.  When such a program has been completed, we shall be able to understand the extent to which our ignorance of the short distance details of the state can be exchanged for a scale-dependent, renormalized initial state and, more importantly, the size of the corrections we can expect to observe.

\begin{acknowledgments}

\noindent 
This work was supported in part by DOE grant DE-FG03-91-ER40682.  I have greatly benefitted from conversations with Rich Holman and Matt Martin.

\end{acknowledgments}

\appendix

\section{Helicity eigenstates}
\label{twospin}

Throughout this article, we work in the Dirac representation for the $\gamma$-matrices,
\begin{equation}
\gamma^0 = \pmatrix{{\bf 1} &0 \cr 0 &-{\bf 1}\cr}
\quad
\vec\gamma = \pmatrix{0 &\vec\sigma \cr -\vec\sigma &0\cr}
\quad
\gamma_5 = \pmatrix{0 &{\bf 1} \cr {\bf 1} &0\cr} . 
\label{Diracrep}
\end{equation}

The two-component positive and negative helicity eigenspinors are denoted respectively by $\varphi^{(s)}_{\hat k}$ and $\chi^{(s)}_{\hat k}$.  Their normalizations are fixed to satisfy
\begin{eqnarray}
\sum_s [\varphi^{(s)}_{\hat k}]^i [\varphi^{(s)\dagger}_{\hat k}]_j &=& \delta^i_j 
\nonumber \\
\sum_s [\chi^{(s)}_{\hat k}]^i [\chi^{(s)\dagger}_{\hat k}]_j &=& \delta^i_j 
\label{spinnorm}
\end{eqnarray}
and 
\begin{eqnarray}
\sum_s s[\varphi^{(s)}_{\hat k}]^i [\varphi^{(s)\dagger}_{\hat k}]_j 
&=& [\hat k\cdot \vec\sigma]^i_j 
\nonumber \\
\sum_s s[\chi^{(s)}_{\hat k}]^i [\chi^{(s)\dagger}_{\hat k}]_j 
&=& - [\hat k\cdot \vec\sigma]^i_j 
\label{Dspinnorm}
\end{eqnarray}
where the indices $i$ and $j$ refer to the components of the spinors.  If in addition we would like these spinors to obey $\varphi^{(s)}_{-\hat k} = \chi^{(s)}_{\hat k}$, then this condition along with Eqs.~(\ref{omegadef}--\ref{spinnorm}) is enough to fix the form of the helicity eigenspinors,
\begin{eqnarray}
\varphi^{(+)}_{\hat k} = - \chi^{(-)}_{\hat k} &=& {e^{-i\delta}\over\sqrt{2}} 
\pmatrix{ {\hat k_x - i\hat k_y\over\sqrt{1-\hat k_z}}\cr
          \sqrt{1-\hat k_z}\cr}
\nonumber \\
\varphi^{(-)}_{\hat k} = \chi^{(+)}_{\hat k} &=& {e^{i\delta}\over\sqrt{2}} 
\pmatrix{ - \sqrt{1-\hat k_z}\cr
          {\hat k_x + i\hat k_y\over\sqrt{1-\hat k_z}}\cr}
\label{Hspinors}
\end{eqnarray}
where
\begin{equation}
e^{2i\delta} = -i 
{\textstyle{\sqrt{\hat k_x^2 + \hat k_y^2}\over \hat k_x + i\hat k_y}} . 
\label{deltadef}
\end{equation}
Note that these two-component spinors satisfy
\begin{eqnarray}
\chi^{(+)}_{\hat k} &=& \varphi^{(+)}_{-\hat k} 
= -i \sigma^2 [\varphi^{(+)}_{\hat k}]^*
\nonumber \\
\chi^{(-)}_{\hat k} &=& \varphi^{(-)}_{-\hat k} 
= -i \sigma^2 [\varphi^{(-)}_{\hat k}]^* . 
\label{Hspinorsrels}
\end{eqnarray}
Later we shall use the following spinor product relation,
\begin{equation}
\Bigl| \varphi^{(+)}_{\hat k_1} \cdot 
\varphi^{(-)\dagger}_{\hat k_2} \Bigr|^2
= {1\over 2} \left( 1 - \hat k_1 \cdot \hat k_2 \right) ,
\label{spinsummed}
\end{equation}
which vanishes when $\hat k_2 = \hat k_1$.

\end{document}